\documentclass[runningheads]{llncs}

\usepackage{amsmath}
\usepackage{amssymb}
\usepackage{CJK}
\usepackage{url}
\usepackage{float}
\usepackage{graphicx}
\usepackage{diagbox}
\usepackage{multirow}
\usepackage{wrapfig}
\usepackage[]{graphbox}
\usepackage{algorithm}
\usepackage{algpseudocode}
\usepackage{booktabs}
\usepackage{xcolor}
\usepackage{subfigure}
\usepackage{tikz} 
\usetikzlibrary{shadows,arrows,shapes}
\pgfdeclarelayer{background}
\pgfdeclarelayer{foreground}
\pgfsetlayers{background,main,foreground}
 
\begin{document}
\begin{CJK*}{UTF8}{gbsn}
\title{Deep Learning for Low-Field to High-Field MR: Image Quality Transfer with \\ Probabilistic Decimation Simulator}
\titlerunning{IQT with PDS}  % abbreviated title (for running head)
\author{Hongxiang Lin\inst{1}\and
Matteo Figini\inst{1}\and
Ryutaro Tanno\inst{1,2} \and
Stefano B. Blumberg\inst{1} \and
Enrico Kaden\inst{1} \and
Godwin Ogbole\inst{3}\and
Biobele J. Brown\inst{4} \and
Felice D'Arco\inst{5} \and\\
David W. Carmichael\inst{6,7} \and
Ikeoluwa Lagunju\inst{4} \and
Helen J. Cross\inst{5,6} \and \\
Delmiro Fernandez-Reyes\inst{1,4} \and 
Daniel C. Alexander\inst{1}}

%\author{Hongxiang Lin\inst{1}\orcidID{0000-0001-6643-327X} \and
% Matteo Figini\inst{1}\orcidID{0000-0002-8238-2262} \and
% Ryutaro Tanno\inst{1,2} \and
% Stefano B. Blumberg\inst{1}\orcidID{0000-0002-8238-2262} \and
% Enrico Kaden\inst{1} \and
% Godwin Ogbole\inst{3}\orcidID{0000-0003-0431-7198} \and
% Biobele J. Brown\inst{4}\orcidID{0000-0001-6981-9872} \and
% Felice D'Arco\inst{5}\orcidID{0000-0001-7396-591X} \and
% David W. Carmichael\inst{6,7}\orcidID{0000-0001-9972-0718} \and
% Ikeoluwa Lagunju\inst{4}\orcidID{0000-0001-9814-1028} \and
% Helen J. Cross\inst{5,6}\orcidID{0000-0001-7345-4829} \and
% Delmiro Fernandez-Reyes\inst{1,4,8}\orcidID{0000-0001-5070-9198} \and 
% Daniel C. Alexander\inst{1}\orcidID{0000-0003-2439-350X}}
% index{Lin, Hongxiang}
% index{Figini, Matteo}
% index{Tanno, Ryutaro}
% index{Blumberg, Stefano B.}
% index{Kaden, Enrico}
% index{Ogbole, Godwin}
% index{Brown, Biobele}
% index{D'Arco, Felice}
% index{Carmichael, David}
% index{Lagunju, Ikeoluwa}
% index{Cross, Helen}
% index{Fernandez-Reyes, Delmiro}
% index{Alexander, Daniel C.}
%
\authorrunning{Hongxiang Lin et al.} % abbreviated author list (for running head)
% \tocauthor{Ivar Ekeland, Roger Temam, Jeffrey Dean, David Grove,
% Craig Chambers, Kim B. Bruce, and Elisa Bertino}
%
\institute{Centre for Medical Image Computing and Department of Computer Science,\\ University College London, UK \and Machine Intelligence and Perception Group, Microsoft Research Cambridge, UK \and Department of Radiology, College of Medicine, University of Ibadan, Nigeria \and Department of Paediatrics, College of Medicine, University of Ibadan, Nigeria \and Great Ormond Street Hospital for Children, London, UK \and UCL Great Ormond Street Institute of Child Health, UK \and Department of Biomedical Engineering, King's College London, UK\\ \email{harry.lin@ucl.ac.uk}} 

\maketitle              % typeset the title of the contribution

\begin{abstract}
%% version three
MR images scanned at low magnetic field ($<1$T) have lower resolution in the slice direction and lower contrast, due to a relatively small signal-to-noise ratio (SNR) than those from high field (typically 1.5T and 3T). We adapt the recent idea of Image Quality Transfer (IQT) to enhance very low-field structural images aiming to estimate the resolution, spatial coverage, and contrast of high-field images. Analogous to many learning-based image enhancement techniques, IQT generates training data from high-field scans alone by simulating low-field images through a pre-defined decimation model. However, the ground truth decimation model is not well-known in practice, and lack of its specification can bias the trained model, aggravating performance on the real low-field scans. In this paper we propose a probabilistic decimation simulator to improve robustness of model training. It is used to generate and augment various low-field images whose parameters are random variables and sampled from an empirical distribution related to tissue-specific SNR on a 0.36T scanner. The probabilistic decimation simulator is model-agnostic, that is, it can be used with any super-resolution networks. Furthermore we propose a variant of U-Net architecture to improve its learning performance. We show promising qualitative results from clinical low-field images confirming the strong efficacy of IQT in an important new application area: epilepsy diagnosis in sub-Saharan Africa where only low-field scanners are normally available. 
\end{abstract}
%%%%%%%%%%%%%%%%%%%%%%%%%
\section{Introduction}
\begin{figure}[t]
    \scriptsize
    \begin{tabular}{cccc}
        \multicolumn{2}{c}{Coronal View} & \multicolumn{2}{c}{Axial View} \\
         \includegraphics[width=.24\textwidth]{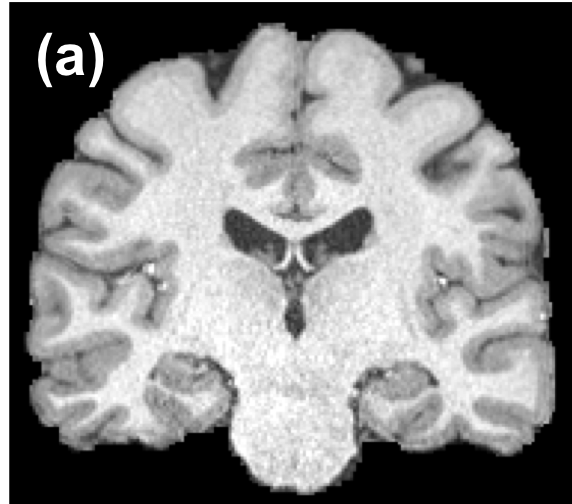} &
         \includegraphics[width=.24\textwidth]{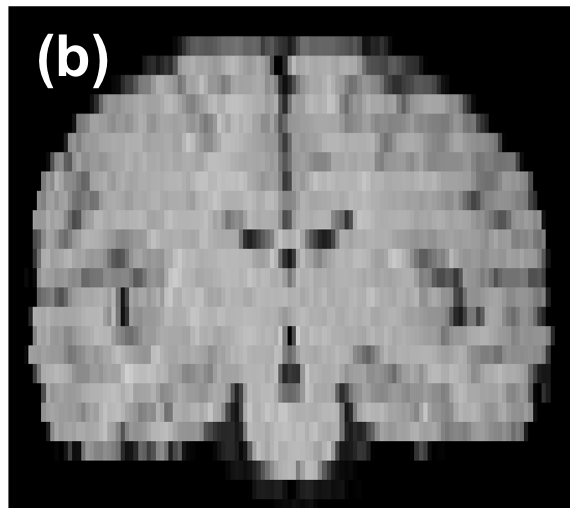} &
         \includegraphics[width=.24\textwidth]{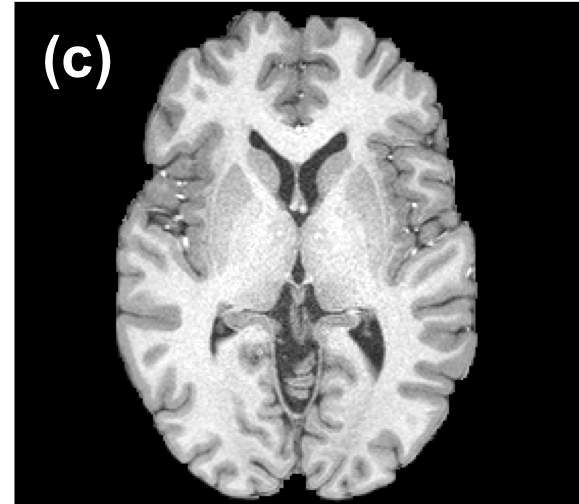} &
         \includegraphics[width=.24\textwidth]{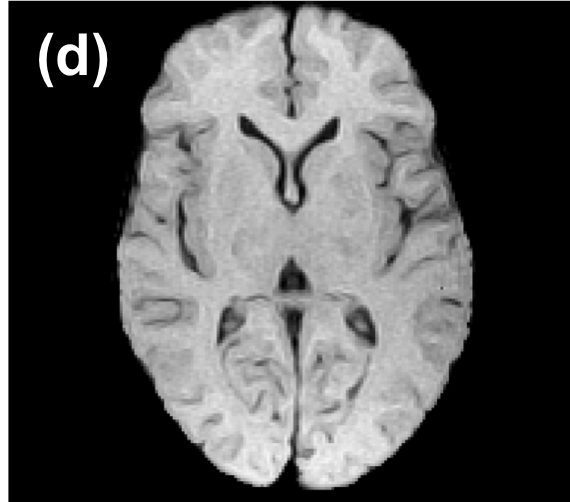} \\
         High-field & Low-field & High-field & Low-field \\
    \end{tabular}
    \caption{High-field vs low-field MR scans: (a-b) Resolution change on coronal plane; (c-d) Contrast change on axial plane. Data sources: (a, c) 3T MRI from Human Connectome Project~\cite{Sotiropoulos2013}; (b, d) 0.36T MRI acquired from University College Hospital, Ibadan.}
    \label{fig:1}
\end{figure}

% Background onf low-field and HF image
Magnetic Resonance Imaging (MRI) is now ubiquitous in neurology with a strong trend towards the use of high-field scanners, with 1.5T and 3T being the current clinical standard. However, low-field MRI scanners, less than 1T, are still common in low and middle income countries (LMICs), due to limited funds and frequent power outages. Low-field scanners suffer from lower signal-to-noise ratio (SNR) than high field at equivalent spatial resolution. To counteract the SNR reduction, practitioners commonly acquire images with non-adjacent thick slices to reduce the acquisition time and cross-talk artifacts in brain MRI scenario~\cite{Wadghiri2001}. This leads to resolution reduction in the slice direction compared with the in-plane resolution and a loss of information due to gaps between slices; see~Fig.~\ref{fig:1}(a-b). Moreover, the contrast between grey matter (GM) and white matter (WM) may be worse than in high field even at equivalent SNR and spatial resolution as illustrated in Fig.~\ref{fig:1}(c-d).

% data harmonisation/image translation
In this study, we aim to learn an image-translation mapping from low field to high field to perform super-resolution and contrast enhancement. In the literature, mathematical models have been proposed to describe the variation of MRI signal with the magnetic field~\cite{Marques2019,brown2014magnetic}, but such models are simplistic and do not include all effects on the final images, such as variability in the acquisition process. Furthermore, the reconstruction of missing information between the acquired slices is severely ill-posed, which hinders the practical capability of producing high-field like images. Several approaches in the literature aim to solve related problems. Bahrami \textit{et al}~\cite{Bahrami2016} proposed a multi-level Canonical Correlation Analysis for estimating 7T from 3T images using paired training data. Wolterink \textit{et al}~\cite{wolterink2017} used the idea of cycle consistency to leverage the abundance of unpaired training sets and learn to synthesise CT from MRI. This approach is, however, known to be susceptible to hallucinations and may introduce spurious features in the output images~\cite{cohen2018distribution}.

% iqt, self-supervision
Image Quality Transfer (IQT) is a machine learning framework used to enhance low-quality clinical data to the abundant neurological information in high-quality images. Most implementations of IQT simulate low quality data from high quality providing matched-paired for training. In~\cite{Alexander2017,tanno2016bayesian,Tanno2017,Blumberg2018} for instance, the corresponding low-field data are synthesised by downsampling and matching voxel-wise intensities coming from prior or empirical knowledge about actual low-field data. However, the trained model strongly depends on the accuracy of low-field synthesis. To improve model generalisability, the prediction of a trained model should be built on unseen test data with less dependency of simulation. 

% data augmentation
In this paper, we build on the IQT framework to construct a mapping that estimates high-field images from the matched low-field inputs. The paired data, particularly in large numbers, are hard to acquire in one area due to the rare availability of high-field scanners in LMICs and low-field scanners in high income countries (HICs). Our key technical contribution is to propose a probabilistic decimation (downsampling) model to improve robustness of IQT training and to enhance images from low-field scanners. More specifically, low-field data generation comes from a probabilistic model which comprises random tissue-specific intensity statistics (e.g. SNR) and probabilistic semantic segmentation. We assume that an \textit{a priori} distribution related to the tissue-specific SNR is available. The segmentation mask estimated by Statistical Parametric Mapping~\cite{Ashburner2005} is also probabilistic in terms of the tissue type. Therefore for one high-field subject, we can simultaneously generate the corresponding multiple low-field data and form the paired training data, a novel way of performing data augmentation. We then learn the low-field-to-high-field transformation by adapting the U-Net architecture~\cite{Heinrich2017} with a super-resolution module, a ``bottleneck block", extending its depth to enable it to capture more global features of image contrast.

\section{Methods}
\subsection{Formulation}\label{sec:2.1}
Let a 3D low-field input patch $x$ of size $w\times h\times d$ be corrupted by smoothing, low contrast, and random noise. It is randomly cropped from the original low-field MR volume denoted by $X$. Our aim is to reconstruct the sub-voxel information in the slice thickness direction and to attain the high SNR and contrast transferring to the corresponding high-field output patch $y$ of size $w\times h\times kd$, where $k$ is an up-sampling rate. Then we assemble all output patches into a high-field MR volume denoted by $Y$. The relationship between $x$ and $y$ is modelled by a degradation process of image quality, described by a function $S$ such that
\begin{equation}\label{iqt_formulation}
    x = S(y, \vec{\alpha})+\epsilon,
\end{equation}
where $\vec{\alpha}$ denotes a vector of SNR components corresponding to prior knowledge of WM and GM in the low-field input volume, i.e. $\vec{\alpha}=(SNR_X^{WM},SNR_X^{GM})$. It is randomly sampled from the Gaussian distribution $\mathcal{N}(\mathbf{\vec{\mu}},\Sigma)$ where $\vec{\mu}$ is a mean vector and $\Sigma$ is a covariance matrix. $\epsilon$ denoting background noise has a Gaussian distribution $\mathcal{N}(0,\sigma_{BG}^2)$. Section~\ref{sec:2.2} will specify the formulation and algorithm for modelling $S$. We then employ deep learning, specifically a convolutional neural network, to estimate the inverse mapping $S^\dagger$.

We use a given $M$-paired training set $\mathcal{T}_M=\{(x_i, y_i)\}_{i=1}^M$ with a fixed $\vec{\alpha}$ to train our convolutional neural networks over all sampled patches from all MR volumes. We optimise the network parameters $\theta$ by minimising the average of the pixel-wise mean squared error (MSE) denoted by $\|\cdot\|_2^2$ over all training sets:
\begin{equation}
    \theta^* = \arg\min_{\theta}\frac{1}{M}\sum_{i=1}^M\|S_\theta^\dagger(x_i)-y_i\|_2^2.
\end{equation}

\subsection{Probabilistic Decimation Simulator}\label{sec:2.2}
% write a new model here.
Equation~\eqref{iqt_formulation} enables us to produce additional training data by randomly sampling the coefficient $\vec{\alpha}$ from an \textit{a priori} distribution, forming the so-called probabilistic decimation simulator. It translates the voxel-wise low-field SNRs, related to the sampled $\vec{\alpha}$ and the tissue category, to the high-field image and down-samples with a factor of $k$. We use this simulator to generate $N$ low-field patches for each high-field patch $y_i$ and form a new training set $\mathcal{T}_{M,N}=\{(x_{ij}, y_i)|i=1,\cdots,M, j=1,\cdots,N\}$. Henceforth, the new model is trained on the augmented set $\mathcal{T}_{M,N}$ with the following expression:
\begin{equation}
    \theta^* = \arg\min_{\theta}\frac{1}{MN}\sum_{i=1}^M\sum_{j=1}^N\|S_\theta^\dagger(x_{ij})-y_i\|_2^2.
\end{equation} 

We develop Alg.~\ref{alg:1} for implementing the probabilistic decimation simulator for neural images. We transform high-field images $Y(\mathbf{v})$ to synthetic low-field images denoted by $\hat{X}(\mathbf{v})$ for any voxel coordinate $\mathbf{v}$ by adapting the SNR in WM and GM to the values obtained in our reference low-field dataset. We assume that SNRs of WM and GM have a 2D Gaussian distribution and the background noise in the low-field or the high-field images has a 1D Gaussian distribution with a zero mean and a standard deviation of $\sigma_{X}$ or $\sigma_{Y}$, respectively. We also assume $\sigma_{X}\gg \sigma_{Y}$ since the random noise in high field is negligible. The simulation procedure starts with the skull-stripped $Y(\mathbf{v})$ with isotropic voxels of length $e_z$. We then down-sample along the slice thickness direction (vertical, or $z$-direction). A $ 1D $ Gaussian filter $h_\sigma(z)=\frac{1}{\sigma\sqrt{2\pi}}{e^{-z^2/(2\sigma^2)}}$ is applied to the high-field images along the $z$-direction, where the $\sigma$ is linked to a full-width at half maximum (FWHM): $\textrm{FWHM}=2\sqrt{2\ln2}\sigma$. The FWHM of the Gaussian filter is set to the slice thickness, or in terms of $\sigma$: $\sigma=k e_z/\sqrt{8\ln 2}$. Then the distance between slices is set to be larger than this slice thickness, emulating the gap between slices. The slices of the original image falling in the gaps have virtually no effect on the signal in the simulated image, similar to what happens in real acquisitions. The high-field images $Y(\mathbf{v})$ are first segmented into tissue categories $j=WM, GM, others$ (denoted by $M^j(\mathbf{v})$) using the unified segmentation algorithm in Statistical Parametric Mapping~\cite{Ashburner2005}. In this algorithm, the mask $M^j(\mathbf{v})$ corresponds to the probability that each voxel $\mathbf{v}$ belongs to the tissue category $j$. SNR of the high-field image with respect to the tissue category $j$ is defined as:
\begin{equation}
   SNR_{Y}^j = \frac{\sum_{\mathbf{v}}M^j(\mathbf{v})Y(\mathbf{v})}{\sigma_{Y}\sum_{\mathbf{v}}M^j(\mathbf{v})}.
\end{equation}
This allows us to evaluate ratios of low-field-to-high-field image intensity for both WM and GM; see Step 5. We then re-scale the high-field images with the ratios of image intensity according to tissue category, which results in the synthetic low-field images $\hat{X}(\mathbf{v})$. We finally add Gaussian white noise to $\hat{X}(\mathbf{v})$, with a standard deviation of $\sigma_{X}$. 
\begin{algorithm}[t]
\caption{Probabilistic Decimation Simulator for low-field Image}
% \hspace*{\algorithmicindent} 
\textbf{Input:} high-field Images $Y(\mathbf{v})$, masks $M^j(\mathbf{v})$ for $j=WM,GM,others$, downsampling scale $k\in\mathbb{N}$, background noise levels $\sigma_{X}$ and $\sigma_{Y}$, low-field SNR distribution $\mathcal{N}(\vec{\mu}, \Sigma)$.
\begin{algorithmic}[1]
\State $Y_{\downarrow k}(\mathbf{v})=Y_{\downarrow k}(\tilde{v}, v')=\sum_{v''}Y(\tilde{v}, kv'-v'')h_{\sigma}(v'')$; \Comment{Downsample on $v''$ component.}
\State $Y_{\downarrow k}^j(\mathbf{v})=M^j(\mathbf{v}) Y_{\downarrow k}(\mathbf{v})$;\Comment{Apply masks.}
\State $SNR_Y^j=\sum_{\mathbf{v}}Y_{\downarrow k}^j(\mathbf{v})/\left(\sigma_{Y}\sum_{\mathbf{v}}M^j(\mathbf{v})\right)$;\Comment{Compute SNRs for high field.}
\State $(SNR_X^{WM}, SNR_{X}^{GM})\sim\mathcal{N}(\vec{\mu}, \Sigma)$;\Comment{Sample SNRs for low field.}
\State $l^j=\left\{\begin{array}{ll}
SNR_X^j/SNR_Y^j, & j=WM,GM, \\
    1, & others; \\
\end{array}\right.$\Comment{Evaluate ratio of image intensity.}
\State $\hat{X}(\mathbf{v})=\sum_{j\in\{WM,GM,others\}}l^j Y_{\downarrow k}^j(\mathbf{v})$; \Comment{Transfer contrast.}
\State $\hat{X}_{\epsilon}(\mathbf{v})=\hat{X}(\mathbf{v})+\epsilon(\mathbf{v})$ where $\epsilon(\mathbf{v})\sim\mathcal{N}(0, \sigma_{X}^2)$. \Comment{Add noise.}
\end{algorithmic}
% \hspace*{\algorithmicindent} 
\textbf{Output:} Noisy synthetic low-field image $\hat{X}_{\sigma}(\mathbf{v})$.
\label{alg:1}
\end{algorithm}

\subsection{Deep Learning Framework}\label{sec2.1}
The classical 3D isotropic U-Net~\cite{Cicek2016} maps two identical-size cubes serving as input and output through the encoder-decoder framework. Each level, defined as a collection of operations in between two shape deformations, for a typical U-Net consists of several convolutional layers together with a pooling layer. The activation from each level in the encoder is concatenated to the input features to the same level in the decoder, enabling the network to integrate both local and global image features. U-Net uses the ``same" zero-padding technique so that feature sizes keep invariant during convolution.

In this work, we extend the U-Net architecture into mapping input and output patches differing with up-scaling factor $k$ in the slice direction. 
% A similar framework was proposed in~\cite{Heinrich2017}. 
Considering the case of $k=4$ illustrated in Fig.~\ref{fig:unet}, this anisotropic U-Net first partially down-samples the first two dimensions until the down-scaling features become isotropic and thereafter conducts isotropic down- and up-sampling. To achieve this, we define the following two operations:

\textbf{Bottleneck Block.} To incorporate a super-resolution transformation into U-Net, we propose a bottleneck block used to connect corresponding levels of the contracting and expanding paths, as shown in Fig.~\ref{fig:unet}(b). The design is inspired by bottleneck block in ResNet~\cite{He2016} and FSRCNN~\cite{Dong2016}. The bottleneck block $BB(b,u)$ has three hyperparameters: the input filter $f$, the number of shrinking layers $b$ and the up-sampling scaling factor $u$. It shrinks half of the filters on consecutive $3\times 3\times 3$ convolutional layers between two endpoint convolutions with a kernel size of $1\times 1\times 1$. All convolution layers are activated by Rectified Linear Unit (ReLU) with Batch Normalization (BN). The skip connection enables the training of deeper networks~\cite{He2016}. Resolution change is efficiently carried out by a transpose convolution, or deconvolution, with the same kernel and stride of $(1,1,u)$.

\textbf{Residual Core.} To have more convolutional layers on each level, the residual core that is a revision of residual element in~\cite{Guerrero2018} is introduced in Fig.~\ref{fig:unet}(c). This is a combination of several sequential $3\times 3\times 3$ convolutional layers, followed by ReLU and BN layers, skip connected with an $1\times 1\times 1$ fully convolutional layer. Then the output is attained before ReLU and BN again. Utilizing the consecutive convolutional layers enlarges each receptive field on each level.
Moreover, the appended skip connection is able to avoid the vanishing gradient problem in neural networks with gradient-based learning methods.
\begin{figure}[t]
    % Define block styles  
    \tikzstyle{materia}=[draw, fill=green!20, text width=6.0em, text centered,
      minimum height=1.5em]
    \tikzstyle{practica} = [materia, text width=2em, rounded corners, drop shadow]
    \tikzstyle{RC} = [practica, fill=yellow!20, text centered]
    \tikzstyle{MP} = [practica, fill=red!20,  text centered]
    \tikzstyle{Deconv} = [practica, fill=purple!20, text centered]
    \tikzstyle{Concat} = [practica, fill=green!20, text centered]
    \tikzstyle{BB} = [practica, text width=2.3em, fill=orange!20, text centered]
    \tikzstyle{Conv} = [practica, fill=blue!20, text centered]
    \tikzstyle{ReLUBN} = [practica, fill=blue!20, text centered]
    \tikzstyle{texto} = [above, text width=18em, text centered]
    \tikzstyle{linepart} = [draw, thick, color=black!50, -latex', densely
    dotted]
    \tikzstyle{line} = [draw, thick, color=black!50, -latex']
    \tikzstyle{ur}=[draw, text centered, minimum height=0.01em]
    \tikzstyle{sum} = [draw, minimum height=1em, minimum width=1em]
    % Define distances for bordering
    \newcommand{\blockdist}{1.3}
    \newcommand{\edgedist}{1.5}
    
    \newcommand{\practica}[2]{node (p#1) [practica]
      {Pr\'actica #1\\{\scriptsize\textit{#2}}}}
    \newcommand{\RC}[3]{node (p#1) [RC, label={[label distance=-0.3em]\tiny\textrm{#2 Ch}}]
      {\tiny\textrm{RC(#3)}}}
    \newcommand{\MP}[2]{node (p#1) [MP, label={-90:\tiny\textrm{#2}}]
      {\tiny\textrm{MP}}}
    \newcommand{\CONV}[3]{node (p#1) [Conv, label={[label distance=-0.3em]\tiny\textrm{#2 Ch}}, label={-90:\tiny\textrm{#3}}]
      {\tiny\textrm{Conv}}}
    \newcommand{\DC}[3]{node (p#1) [Deconv, label={[label distance=-0.3em]90:\tiny\textrm{#2 Ch}} , label={-90:\tiny\textrm{#3}}]
      {\tiny\textrm{Deconv}}}
    \newcommand{\BB}[4]{node (p#1) [BB, label={[label distance=-0.3em]\tiny\textrm{#2 Ch}}]
      {\tiny\textrm{BB(#3,#4)}}}
    \newcommand{\RELUBN}[1]{node (p#1) [ReLUBN, rotate=90, text width=5.5em]
      {\tiny\textrm{ReLU$+$BN}}}
    \newcommand{\CONVRELUBN}[3]{node (p#1) [ReLUBN, rotate=90, text width=5.5em, label={[label distance=0em, text centered, xshift=-1.2em, yshift=0.2em]right:\tiny\textrm{#2 Ch}}, label={[label distance=0em, text centered, xshift=1.2em, yshift=-0.2em]left:\tiny\textrm{#3}}]
      {\tiny\textrm{Conv$+$ReLU$+BN$}}}
    \newcommand{\Concat}[2]{node (p#1) [Concat, label={[label distance=-0.3em]\tiny\textrm{#2 Ch}}]
      {\tiny\textrm{Concat}}}
    \newcommand{\SUMA}[1]{node (p#1) [sum,circle]
      {\scriptsize$+$}}
    % Draw background
    \newcommand{\bg}[5]{%
      \begin{pgfonlayer}{background}
        % Left-top corner of the background rectangle
        \path (#1.north |- #2.east)+(-0,0.7) node (a1) {};
        % Right-bottom corner of the background rectanle
        \path (#3.south |- #4.west)+(+0,-0.8) node (a2) {};
        % Draw the background
        \path[fill=orange!20,rounded corners, draw=black!50, dashed]
          (a1) rectangle (a2);
        \path (a1.east |- a1.south)+(2.8,-0.3) node (u1)[texto]
          {\scriptsize\textit{#5}};
      \end{pgfonlayer}}
    
    % Draw background
    \newcommand{\bgbg}[5]{%
      \begin{pgfonlayer}{background}
        % Left-top corner of the background rectangle
        \path (#1.north |- #2.east)+(-0,0.7) node (a1) {};
        % Right-bottom corner of the background rectanle
        \path (#3.south |- #4.west)+(+0,-0.25) node (a2) {};
        % Draw the background
        \path[fill=yellow!20,rounded corners, draw=black!50, dashed]
          (a1) rectangle (a2);
        \path (a1.east |- a1.south)+(2.5,-0.3) node (u1)[texto]
          {\scriptsize\textit{#5}};
      \end{pgfonlayer}}
      
      % Draw background
    \newcommand{\bgbgbg}[5]{%
      \begin{pgfonlayer}{background}
        % Left-top corner of the background rectangle
        \path (#1.north |- #2.east)+(-0.2,0.6) node (a1) {};
        % Right-bottom corner of the background rectanle
        \path (#3.south |- #4.west)+(+0.2,-0.2) node (a2) {};
        % Draw the background
        \path[fill=gray!20,rounded corners, draw=black!50, dashed]
          (a1) rectangle (a2);
        \path (a1.east |- a1.south)+(0.5,-0.3) node (u1)[texto]
          {\scriptsize\textbf{#5}};
      \end{pgfonlayer}}
    \newcommand{\notebg}[4]{%
      \begin{pgfonlayer}{background}
        % Left-top corner of the background rectangle
        \path (#1.west |- #2.north)+(-0.2,0.1) node (a1) {};
        % Right-bottom corner of the background rectanle
        \path (#3.east |- #4.south)+(+2,-0.2) node (a2) {};
        % Draw the background
        \path[fill=white!20,rounded corners, draw=black!100, dashed]
          (a1) rectangle (a2);
      \end{pgfonlayer}}
    \centering%
    \centering%
    
\begin{tabular}{cc}

\multicolumn{2}{c}{\begin{tikzpicture}[scale=0.75,transform shape]
    % Draw diagram elements
    \path node (p0) [text centered] {\scriptsize{Input data}};
    \path (p0.south)+(0,-0.7) \RC{1}{16}{3};
    \path (p1.south)+(0,-0.7) \MP{2}{2x2x1};
    \path (p2.east)+(0.5,0) \RC{3}{32}{3};
    \path (p3.south)+(0,-0.7) \MP{4}{2x2x1};
    \path (p4.east)+(0.5,0) \RC{5}{64}{3};
    \path (p5.south)+(0,-0.7) \MP{6}{2x2x2};
    \path (p6.east)+(0.5,0) \RC{7}{128}{3};
    \path (p7.south)+(0,-0.7) \MP{8}{2x2x2};
    \path (p8.east)+(0.5,0) \RC{9}{256}{3};
    \path (p9.east)+(0.5,0) \DC{10}{128}{2x2x2};
    \path (p10.north)+(0,0.7) \Concat{11}{256};
    \path (p11.east)+(0.5,0) \RC{12}{128}{3};
    \path (p12.east)+(0.5,0) \DC{13}{64}{2x2x2};
    \path (p13.north)+(0,0.7) \Concat{14}{128};
    \path (p14.east)+(0.5,0) \RC{15}{64}{3};
    \path (p15.east)+(0.5,0) \DC{16}{32}{2x2x2};
    \path (p16.north)+(0,0.7) \Concat{17}{64};
    \path (p17.east)+(0.5,0) \RC{18}{32}{3};
    \path (p18.east)+(0.5,0) \DC{19}{16}{2x2x2};
    \path (p19.north)+(0,0.7) \Concat{20}{32};
    \path (p20.east)+(0.5,0) \RC{21}{16}{3};
    \path (p21.east)+(0.5,0) \CONV{22}{1}{1x1x1};
    \path (p22.north)+(0,0.7) node (p00) [text centered] {\scriptsize{Output data}};
    \path (p1.east)+(5,0) \BB{23}{16}{2}{4};
    \path (p3.east)+(4,0) \BB{24}{32}{2}{2};
    
    % notes
    \path (p13.east)+(1,0) node (p25) [RC,label={0:\tiny\textrm{Residual Core}}] {\tiny\textrm{RC($b$)}};
    \path (p25.south)+(0,-.3) node (p26) [MP,label={0:\tiny\textrm{Max Pooling}}] {\tiny\textrm{MP}};
    \path (p26.south)+(0,-.3) node (p27) [Concat,label={0:\tiny\textrm{Concatenation}}] {\tiny\textrm{Concat}};
    \path (p25.east)+(2.2,0) node (p28) [Deconv,label={0:\tiny\textrm{Deconvolution}}] {\tiny\textrm{Deconv}};
    \path (p28.south)+(0,-.3) node (p29) [BB,label={0:\tiny\textrm{Bottleneck Block}}] {\tiny\textrm{BB($b$,$u$)}};
    \path (p29.south)+(0,-.3) node (p30) [Conv,label={0:\tiny\textrm{Convolution}}] {\tiny\textrm{Conv}};
    % bg
    \notebg{p25}{p25}{p30}{p30}
    
    % Draw line between elements
    \path [line] (p0.south) -- node [left, midway] {\tiny\color{red}{$(32, 32, 8)$}} (p1);
    \path [line] (p1.south) -- node [left, midway] {\tiny\color{red}{$(32, 32, 8)$}} (p2);
    \path [line] (p3.south) -- node [left, near end] {\tiny\color{red}{$(16, 16, 8)$}} (p4);
    \path [line] (p5.south) -- node [left, near end] {\tiny\color{red}{$(8, 8, 8)$}} (p6);
    \path [line] (p7.south) -- node [left, near end] {\tiny\color{red}{$(4, 4, 4)$}} (p8);
    
    % \path (p9.south)+(0, -0.3) node [text centered] {\tiny\color{red}{$(2, 2, 2)$}} (ptext);
    \path (p9.south)+(0, -0.3) node [text centered] {\tiny\color{red}{$(2, 2, 2)$}} ;
        
    \path [line] (p10.north) -- node [right] {\tiny\color{red}{$(4, 4, 4)$}} (p11);
    \path [line] (p13.north) -- node [right] {\tiny\color{red}{$(8, 8, 8)$}} (p14);
    \path [line] (p16.north) -- node [right] {\tiny\color{red}{$(16, 16, 16)$}} (p17);
    \path [line] (p19.north) -- node [right] {\tiny\color{red}{$(32, 32, 32)$}} (p20);
    \path [line] (p22.north) -- node [right] {\tiny\color{red}{$(32, 32, 32)$}} (p00);
    
    \path [line] (p1.east) -- node [above,midway] {\tiny\color{red}{$(32, 32, 8)$}} (p23);
    \path [line] (p23.east) -- node [above,midway] {\tiny\color{red}{$(32, 32, 32)$}} (p20);
    \path [line] (p3.east) -- node [above,midway] {\tiny\color{red}{$(16, 16, 8)$}} (p24);
    \path [line] (p24.east) -- node [above,midway] {\tiny\color{red}{$(16, 16, 16)$}} (p17);
    \path [line] (p5.east) -- node [above,midway] {\tiny\color{red}{$(8, 8, 8)$}} (p14);
    \path [line] (p7.east) -- node [above,midway] {\tiny\color{red}{$(4, 4, 4)$}} (p11);
\end{tikzpicture}}\\  
\multicolumn{2}{c}{(a) Anisotropic U-Net Architecture.}\\
\begin{tikzpicture}[scale=0.7,transform shape]
    \path node (p0) [text centered,rotate=90] {\scriptsize{Input}};
    \path (p0.south)+(.5,0) \CONVRELUBN{1}{$f/2$}{1x1x1};
    \path (p1.south)+(.7,0) \CONVRELUBN{3}{$f/2$}{3x3x3};
    \path (p3.south)+(.2,0) node (p5) [rotate=90] {$\cdots$};
    \path (p5.south)+(.2,0) \CONVRELUBN{6}{$f/2$}{3x3x3};
    \path (p6.south)+(.7,0) \CONVRELUBN{8}{$f$}{1x1x1};
    \path (p8.south)+(.7,0) \SUMA{10};
    \path (p10.east)+(.7,0) node (p11) [Deconv, rotate=90, label={[text centered, xshift=-1.2em, yshift=0.2em]right:\tiny\textrm{$f$ Ch}} , label={[text centered, xshift=1.2em, yshift=-0.2em]left:\tiny\textrm{1x1x$u$}}] {\tiny\textrm{Deconv}};
    \path (p11.south)+(.7,0) node (p12) [rotate=90] {\scriptsize{Output}};
    
    % Draw line between elements
    \path [line] (p0.south) -- node [above, midway] {} (p1);
    \path [line] (p1.south) -- node [above] {} (p3);
    \path [line] (p6.south) -- node [above] {} (p8);
    \path [line] (p8.south) -- node [left] {} (p10);
    \path [line] (p10.east) -- node [above] {} (p11);
    \path [line] (p11.south) -- node [above] {} (p12);
    \path [line] (p0.west) --+ (0, -1.0)--+(4.4, -1.0)--node [below] {} (p10);
    \bg{p0}{p1}{p12}{p1}{}
    \bgbgbg{p3}{p3}{p6}{p6}{$b$ copies}
\end{tikzpicture} & 
\begin{tikzpicture}[scale=0.7,transform shape]
    \path node (p0) [text centered,rotate=90] {\scriptsize{Input}};
    \path (p0.south)+(.7,0) \CONVRELUBN{3}{$f$}{3x3x3};
    \path (p3.south)+(.2,0) node (p5) [rotate=90] {$\cdots$};
    \path (p5.south)+(.2,0) \CONVRELUBN{6}{$f$}{3x3x3};
    \path (p6.south)+(.7,0) \SUMA{10};
    \path (p10.east)+(.7,0) \RELUBN{11};
    \path (p11.south)+(1,0) node (p12) [rotate=90] {\scriptsize{Output}};
    \path (p5.west)+(0,-1.5) \CONV{13}{$f$}{1x1x1};
    
    % Draw line between elements
    \path [line] (p0.south) -- node [above] {} (p3);
    \path [line] (p6.south) -- node [left] {} (p10);
    \path [line] (p10.east) -- node [above] {} (p11);
    \path [line] (p11.south) -- node [above] {} (p12);
    \path [line] (p0.west) --+ (0, -1.4)--node [left] {} (p13);
    \path [line] (p13.east) --+ (0.9, 0) -- node [below] {} (p10);
    % Draw background
    \bgbg{p0}{p3}{p12}{p13}{}
    \bgbgbg{p3}{p3}{p6}{p6}{$b$ copies}
\end{tikzpicture}\\
(b) Bottleneck Block $BB(b,u)$ & (c) Residual Core $RC(b)$\\
\end{tabular}
\caption{(a) The diagram of anisotropic U-Net (example for the up-scaling factor of $k=4$). The operations, (b) Bottleneck Block $BB(b,u)$ with $f$ filters and (c) Residual Core $RC(b)$ with $f$ filters, are detailed. The round boxes correspond to the different operations illustrated in the bottom right of (a). The number of output channels, abbreviated as “Ch”, and the kernel size are denoted on top and bottom of the boxes. The arrows represent transfer of data with its corresponding shape highlighted.}
\label{fig:unet}
\end{figure}

\section{Experiments}
\subsection{Implementation Details}
\textbf{Datasets.} High-resolution axial T1-weighted images were obtained from the publicly available Human Connectome Project (HCP) dataset~\cite{Sotiropoulos2013}, acquired on a 3T Siemens Connectome scanner with an isotropic voxel size of $0.7 \times 0.7 \times 0.7$ mm$^3$. To investigate sensitivity of the proposed U-Net, we trained it on two training sets with two up-scaling factors of $k=4$ or $8$. Specifically, the slice thickness/gap is $2.1$ mm/$0.7$ mm for $k=4$, and $4.2$ mm/$1.4$ mm for $k=8$. As a reference for low field, T1-weighted images were acquired on a 0.36T MagSense 360 MRI System scanner with a non-isotropic voxel size of $0.9 \times 0.9 \times 7.2$ mm$^3$ including $6.0$ mm slice thickness and $1.2$ mm gaps. The distribution of white matter and grey matter SNRs in the low field was acquired from $28$ image data from children with epilepsy in University College Hospital, Ibadan, whose ages are within a range from 2 to 15 years.

\textbf{IQT Pipeline.}  
In the training stage, we randomly selected $30$ subjects with skull-stripping from HCP dataset and employed them to synthesise the low-field images using Alg.~\ref{alg:1} based on \textit{a priori} variable SNRs. Regarding patch extraction, we cropped the low-field patches with the step size of $8$, $16$, and $16/k$ along $x$-, $y$-, and $z$-directions, respectively. We also cropped the high-field patches with the same volume and position as the corresponding low-field patches. The low-field and high-field patch sizes were $32\times 32\times (32/k)$ and $32\times 32\times 32$, respectively. Then the patches capturing $80\%$ background voxels were excluded from a patch library. 

We examined if overfitting occurred with a validation set and judged the performance of the trained neural network with an evaluation set. We split all $30$ subjects into $12$, $3$, and $15$ for training, validation, and evaluation sets. Moreover, we investigated the image quality by calculating the peak signal-to-noise ratio (PSNR) and the structural similarity index (SSIM)~\cite{wang2004image}. We employed a two-tailed Wilcoxon signed-rank test to determine the statistical significance of the performance difference between two comparing methods.

\textbf{Neural Networks.} We conducted an ablation study on the proposed U-Net, denoted by ANISO U-Net($b$), in the case of $b=2$ or $3$ for shrinking layers in the bottleneck block. We evaluated our networks against the 3D cubic B-spline interpolation and several existing U-Net baselines equivalently switching off the corresponding blocks, i.e. bottleneck block and residual core, in ANISO U-Net. One is an isotropic 3D U-Net (ISO U-Net)~\cite{Cicek2016} implemented with $5$ levels and $3$ convolutional layers per level. The input of ISO U-Net is isotropically interpolated using cubic B-splines. The other one is 3D-SRU-Net~\cite{Heinrich2017} that up-samples each level output on the contraction path before concatenation. It contained $3$ levels for the down-sampling scale $k=4$ and $4$ levels for $k=8$. We unified hyper-parameters of the three U-Nets as follows. Number of filters on the first level was $16$ with the number of filters doubling at each subsequent level. All U-Nets were implemented in Python using Keras library~\cite{chollet2015keras} with Tensorflow backend. They were calculated on a Nvidia GTX 1080 Ti GPU. Training used ADAM~\cite{kingma2014adam} as the optimizer with a starting learning rate of $10^{-3}$ and a decay of $10^{-6}$. We initialized the parameters with Glorot normal initializer~\cite{glorot2010understanding}. The batchsize was $32$ and the loss function is the pixel-wise mean squared error (MSE). All the experiments started converging after about $30$ epochs and we employed early stopping after $5$ epochs of no improvement on the validation set.

\subsection{Evaluation on Fixed SNR Data Sets}\label{sec:3.2}

\begin{figure}[t]
    %\vspace{-.7cm}
    \scriptsize
    \centering
    \begin{tabular}{>{\centering\arraybackslash}>{\centering\arraybackslash}p{.15cm}>{\centering\arraybackslash}p{1.55cm}>{\centering\arraybackslash}p{1.55cm}>{\centering\arraybackslash}p{1.55cm}>{\centering\arraybackslash}p{1.55cm}>{\centering\arraybackslash}p{1.55cm}>{\centering\arraybackslash}p{1.55cm}>{\centering\arraybackslash}p{1.55cm}}
     & Input & Cubic & ISO & 3D-SRU-Net & ANISO & ANISO & Ground \\
     & & & U-Net & & U-Net(2) & U-Net(3) &  Truth  \\
    \rotatebox{90}{$\quad\ $Coronal} &\includegraphics[width=.95\textwidth]{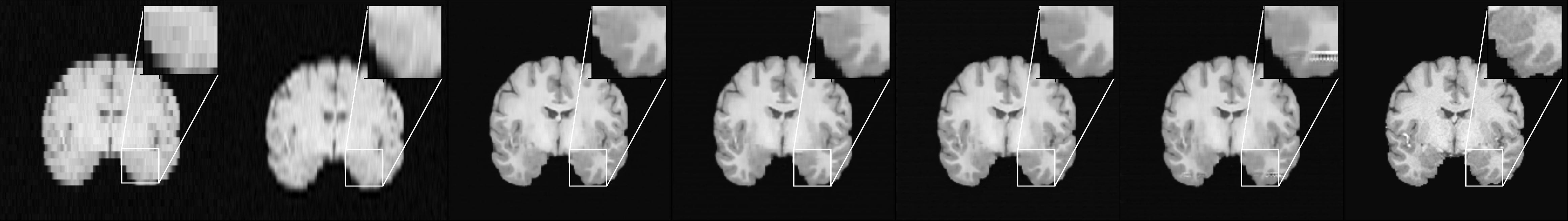}\\
    \rotatebox{90}{$\quad$Sagittal}& \includegraphics[width=.95\textwidth]{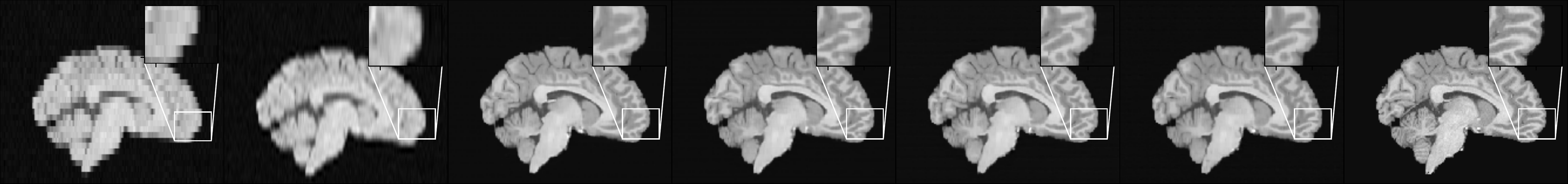}\\
    \end{tabular}
    % \vspace{-2mm}
    \caption{Visualization of U-Net reconstructions with the up-scaling factor $k=8$. }
    \label{fig:4}
    % \vspace{-3mm}
\end{figure}

We evaluated the ability of the proposed U-Net in an ideal case where the SNR-related coefficient $\vec{\alpha}=(SNR_X^{WM},SNR_X^{GM})$ in Eq.~\eqref{iqt_formulation} is deterministic. We fixed the $SNR_X^{WM}$ and $SNR_X^{GM}$ as $61$ and $53$, respectively, in the IQT pipeline by reconstructing images in the evaluation set at Step 4 in Algorithm~\ref{alg:1}. Table~\ref{tab:1} shows that our model, ANISO U-Net(2), achieved the best performance in terms of the average PSNR and SSIM, and especially, significantly outperformed the others in terms of PSNR at $k=4$ and the mean SSIM (MSSIM) at $k=8$. The reconstruction degraded as the up-scaling factor increased. Figure~\ref{fig:4} shows the U-Net reconstructions on coronal and sagittal planes. Qualitatively we observed clear recovery of high resolution information and enhancement of contrast. The reconstructed images from all networks nicely highlighted features visible in the ground truth images that were obscured in the low quality input. The quantitative results in Table~\ref{tab:1} show little difference among the U-Net outputs but they might not be able to reflect subtle qualitative differences. The zoomed patches in Fig.~\ref{fig:4} highlight differences more clearly and we believe ANISO U-Net(2) approximates the ground truth most closely and with the least artefacts as shown in the ANISO U-Net(3) result of Fig.~\ref{fig:4}. Delicately selecting hyper-parameters can avoid overfitting, and hence can mitigate the artifacts. 
\begin{table}[t]
\footnotesize
\centering
\caption{The performance of the proposed model on up-scaling factors $k=4$ or $8$. The mean and standard deviation of PSNR and the mean SSIM (MSSIM) are calculated over $15$ evaluation subjects. For each case, we show the best performance over an ensemble of $5$ trained models. Bold font denotes the best mean or standard deviation. The asterisk $^*$ denotes $p$-value$<0.01$ compared with the rest methods.}
\begin{tabular}{|c|c|c|c|c|}
\hline%\noalign{\smallskip}
%\noalign{\smallskip}
\multirow{2}{*}{Method} & \multicolumn{2}{c|}{$k=4$} & \multicolumn{2}{c|}{$k=8$} \\
                        & PSNR (dB) & MSSIM & PSNR (dB) & MSSIM \\
\hline
%\noalign{\smallskip}
Cubic B-spline & $20.689\pm2.540^*$ & $0.692\pm0.0384^*$ & $18.974\pm2.535^*$ & $0.567\pm0.0471^*$ \\
ISO U-Net      & $30.798\pm2.573$ & $0.916\pm0.0227^*$ & $27.073\pm2.469$ & $0.846\pm0.0278$ \\
3D-SRU-Net     & $30.764\pm2.638$ & $0.922\pm0.0191$ & $27.275\pm2.542$ & $0.847\pm0.0290$ \\
ANISO U-Net(2) & $\mathbf{31.045\pm2.654}^*$ & $\mathbf{0.923\pm0.0197}$ & $\mathbf{27.346\pm2.517}$ & $\mathbf{0.852\pm0.0280}^*$ \\
ANISO U-Net(3) & $30.918\pm2.639$ & $0.921\pm0.0199^*$ & $27.054\pm2.544$ & $0.847\pm0.0282$ \\
\hline
\end{tabular}
\label{tab:1}
\end{table}

\subsection{Evaluation on Variable SNR Data Sets}
We evaluated the performance of several deep learning architectures including the proposed anisotropic U-Net with variable-SNR low-field data. $SNR_X^{WM}$ and $SNR_X^{GM}$ are now sampled from a two-dimensional Gaussian distribution $\mathcal{N}(\mu, \Sigma)$ where the coefficients are:
\begin{equation*}
    % \mu = \left(\begin{array}{l}64.50\\54.14\end{array}\right)
    \mu = (64.50, 54.14)\quad \Sigma = \left(\begin{array}{cc}78.47, & 71.50 \\71.50, & 73.91 \end{array} \right).
\end{equation*}
The simulator shown in Alg.~\ref{alg:1} randomly generated $N$ low-field input images with different SNR for the chosen $15$ training subjects in the HCP data set. We trained the deep learning models on the dataset with the augmenting factor $N=1, 2, 4$ and $8$. We randomly selected $12.5\%$ patches for training in each overlap patch library. For each neural network, an ensemble of $5$ models were trained in terms of different augmented dataset. 

Table~\ref{tab:2} shows the mean and standard deviation of PSNR and MSSIM over $15$ test subjects in terms of the augmented datasets and deep learning architectures. As a result, probabilistic decimation model was generally able to produce more stable reconstruction than the deterministic model if the unseen test data were also generated from the variable SNR. Both accuracy and robustness corresponding to mean and standard deviation of MSSIM improved in various degree as the number of generated low-field image samples increased, and in addition, the performances for the two methods were statistically significant in terms of $N=8$ at $k=8$. Regarding PSNR, the performance upgraded after augmentation but the robustness reflected by the standard deviation did not improve correspondingly. In addition, we observed that PSNR and MSSIM at $k=4$ only slightly improve when the augmenting factor $N$ became larger, which means the improvement of performance arising from augmentation gradually reached an upper bound.
\begin{table}[h]
\footnotesize
\centering
\caption{The performance of probabilistic decimation simulation for augmentation with a factor of $N$. The mean and standard deviation of PSNR and MSSIM are calculated over $15$ evaluation subjects. We show the best performance over an ensemble of $5$ trained models. The ``const" at $N$ samples/subject column means that the models were trained on the fixed SNR data sets as described in Section~\ref{sec:3.2}. Bold font denotes the best mean or standard deviation. The asterisk $^*$ denotes $p$-value$<0.01$ compared with the other augmentation factors.}
\begin{tabular}{|c|c|c|c|c|c|}
\hline%\noalign{\smallskip}
%\noalign{\smallskip}
\multirow{2}{*}{Method} & $N$ samples  & \multicolumn{2}{c|}{$k=4$} & \multicolumn{2}{c|}{$k=8$} \\
                        & $/$subject   & PSNR (dB) & MSSIM & PSNR (dB) & MSSIM \\
\hline
%\noalign{\smallskip}
\multirow{5}{*}{\shortstack{3D-SR\\U-Net}} & ``const" & $27.214\pm3.030$ & $0.871\pm0.0332$ & $24.687\pm\mathbf{2.464}$ & $0.758\pm 0.0371$ \\
               & $1$ & $27.988\pm\mathbf{2.445}$ & $0.861\pm0.0286$ & $23.777\pm2.693$ & $0.757\pm0.0401$ \\
               & $2$ & $29.453\pm2.585$ & $0.901\pm0.0240^*$ & $25.513\pm2.711^*$ & $0.799\pm0.0366^*$ \\
               & $4$ & $\mathbf{30.257}\pm2.647$ & $\mathbf{0.918\pm0.0201}$ & $26.025\pm2.587$ & $0.816\pm0.0339^*$ \\
               & $8$ & $29.958\pm2.541$ & $0.911\pm0.0203$ & $\mathbf{26.391}\pm2.621$ & $\mathbf{0.832\pm0.0316}^*$ \\\hline
\multirow{5}{*}{\shortstack{ANISO\\U-Net(2)}}& ``const" & $27.311\pm3.522$ & $0.870\pm0.0338$ & $24.754\pm2.367$ & $0.769\pm0.0341$ \\
               & $1$ & $28.664\pm2.552$ & $0.890\pm0.0240$ & $23.418\pm\mathbf{2.322}$ & $0.757\pm0.0345^*$ \\
               & $2$ & $29.216\pm\mathbf{2.308}$ & $0.893\pm0.0248$ & $25.862\pm2.617^*$ & $0.803\pm0.0343$ \\
               & $4$ & $30.248\pm2.565$ & $\mathbf{0.916}\pm0.0195$ & $26.231\pm2.613$ & $0.807\pm0.0381$ \\
               & $8$ & $\mathbf{30.344}\pm2.421$ & $0.914\pm\mathbf{0.0188}$ & $\mathbf{27.053}\pm2.398$ & $\mathbf{0.843\pm0.0330}^*$ \\
\hline
\end{tabular}
\label{tab:2}
\end{table}

\subsection{Test on Patient Data}

\begin{figure}[h]
    \scriptsize
    \centering
    \begin{tabular}{cccccc}
    \multicolumn{3}{c}{Axial} & \multicolumn{3}{c}{Coronal}\\
    \includegraphics[height=.12\textheight]{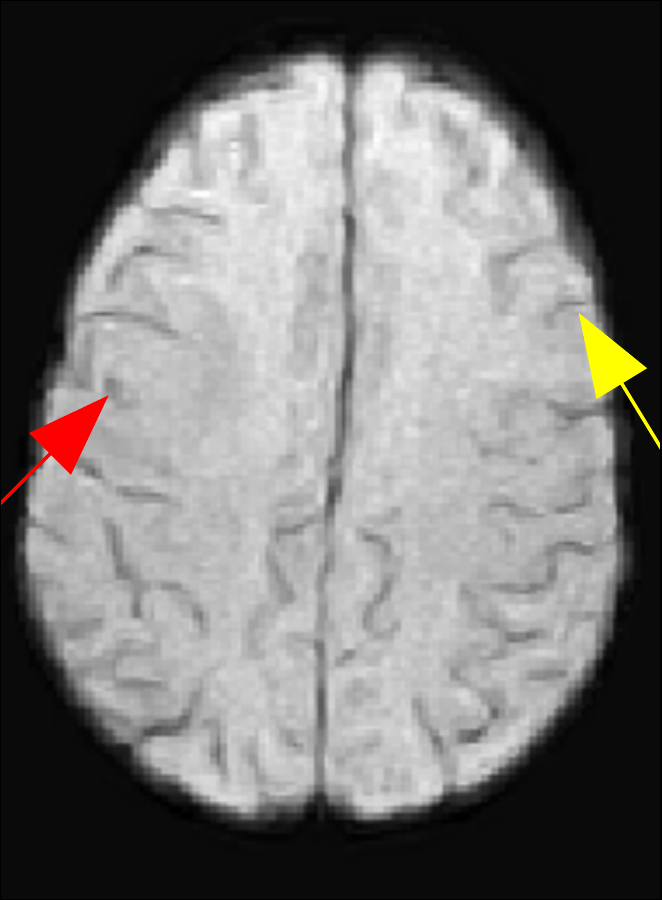}\hspace{-0.5em} & \includegraphics[height=.12\textheight]{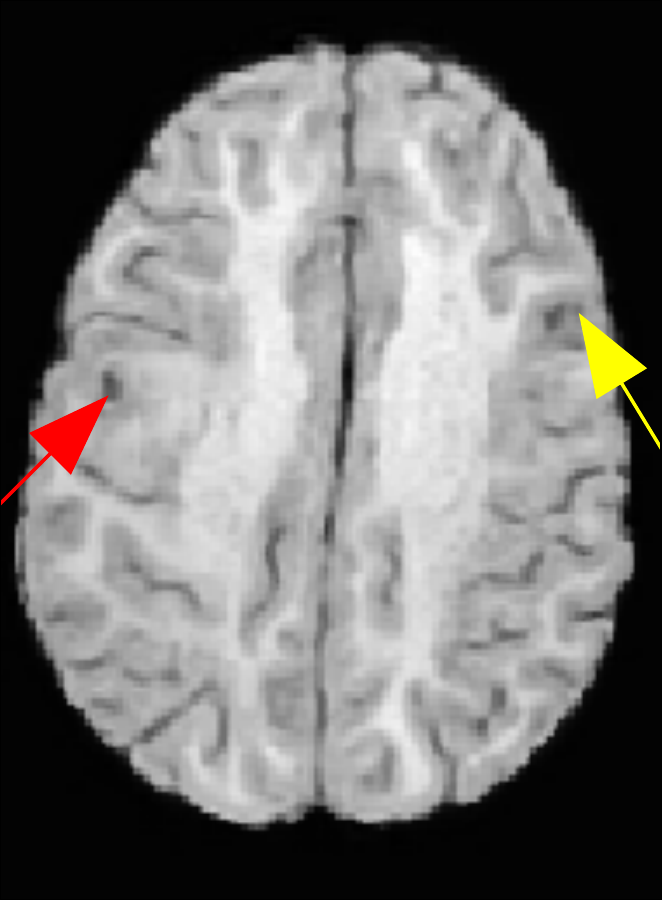}\hspace{-.5em} &
    \includegraphics[height=.12\textheight]{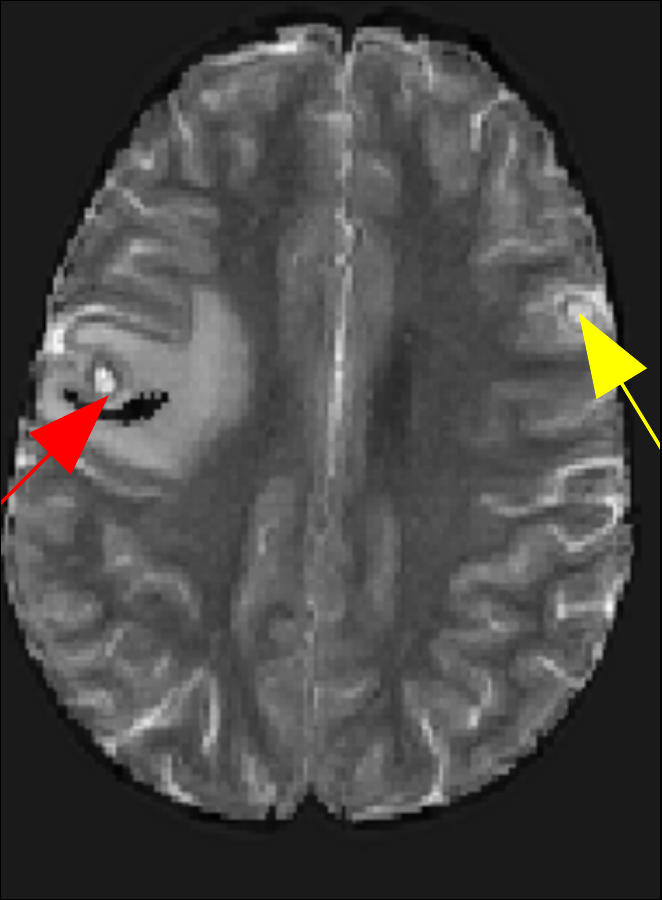}\hspace{1em} &
    \includegraphics[width=.18\textwidth,height=.12\textheight]{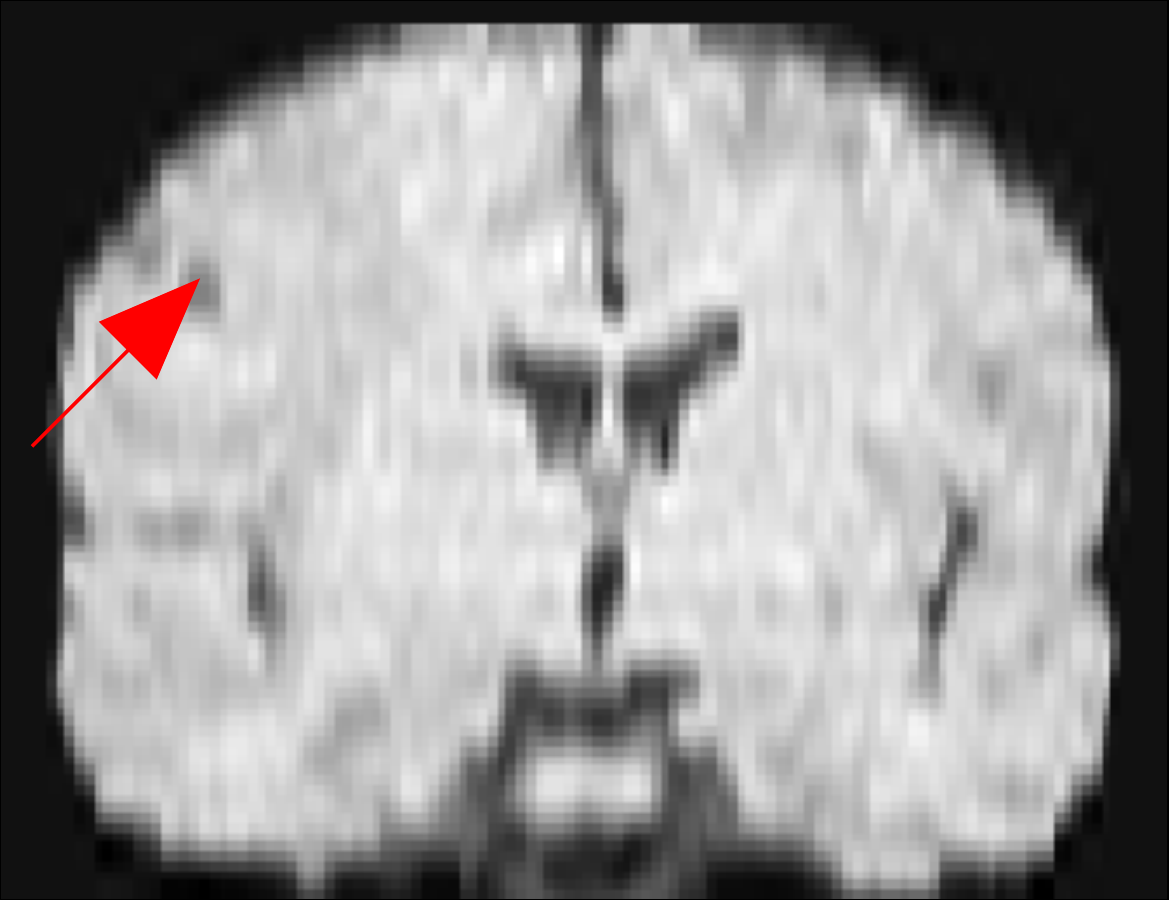}\hspace{-.5em} & \includegraphics[width=.18\textwidth,height=.12\textheight]{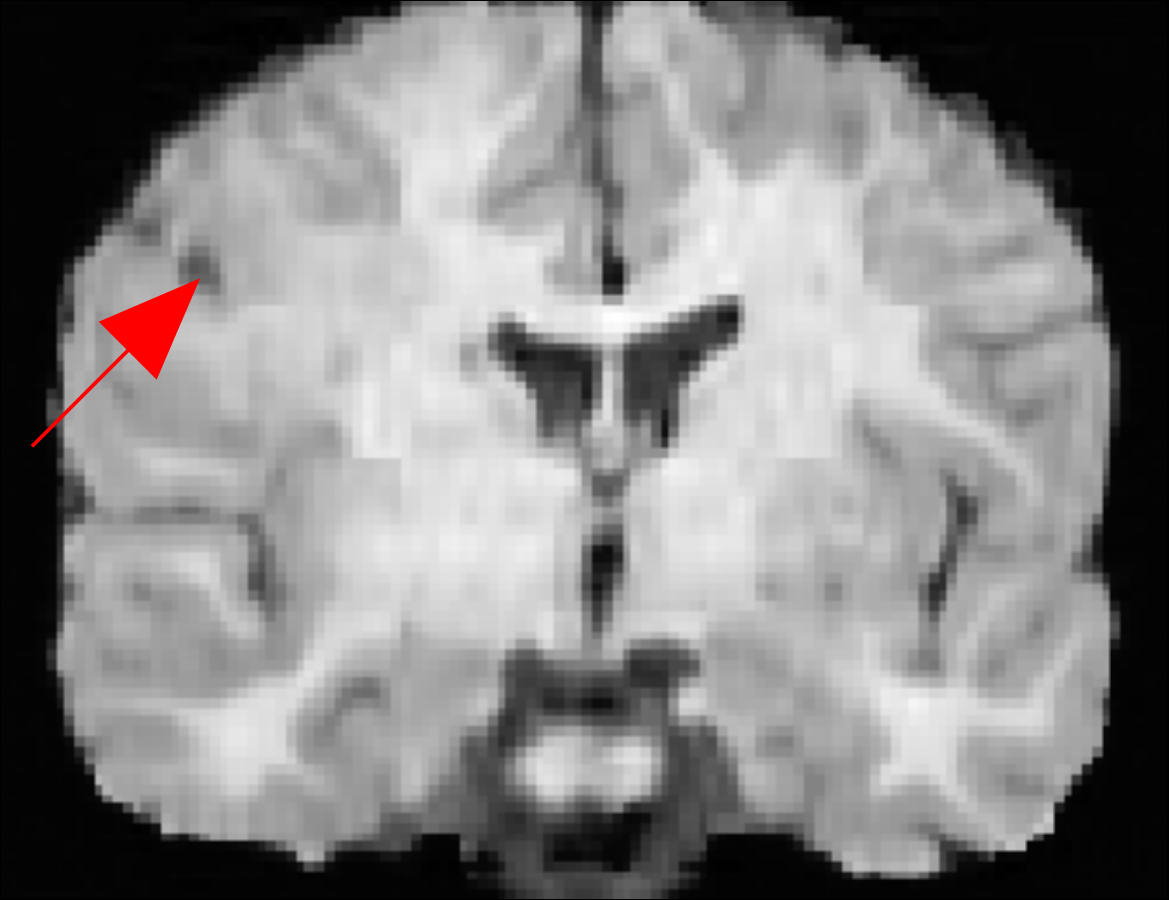}\hspace{-.5em} &
    \includegraphics[width=.18\textwidth,height=.12\textheight]{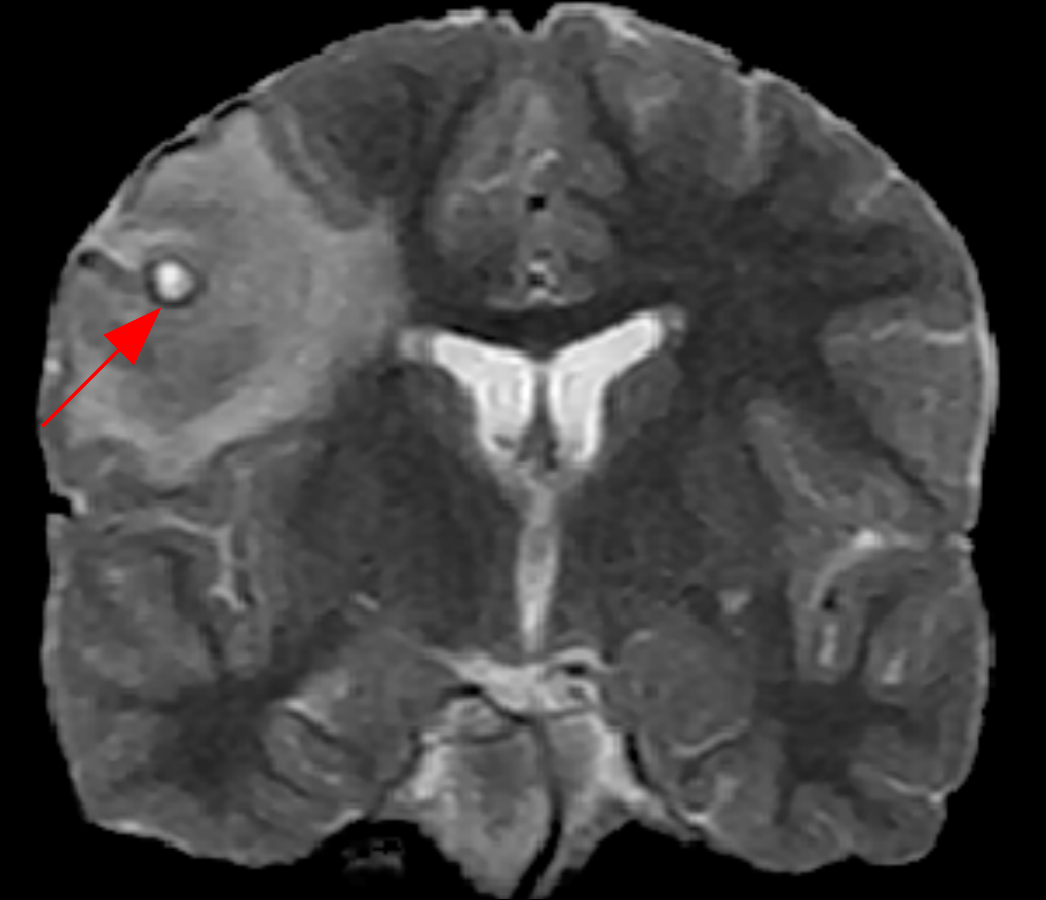}\\
    (a) Input & (b) Enhanced & (c) Reference & (d) Input & (e) Enhanced & (f) Reference \\
    \end{tabular}
    \caption{The IQT prediction on the low-field epileptic patient data for (a-c) axial plane and (d-f) coronal plane. (a) and (d): Low-field T1-weighted input with cubic B-spline interpolation; (b) and (e): IQT-enhanced T1-weighted output using ANISO U-Net(2); (c) and (f): low-field T2-weighted image as a reference of ground truth. Two sub-centimeter parenchymal cystic lesions at the GM-WM junction of the parietal lobes are pointed out by the red and the yellow arrows. They are barely visible in (a) and (d) but greatly enhanced in (b) and (e). (c) and (f), not involved in the IQT experiment, verified their location in an independent acquisition.}
    \label{fig:5}
\end{figure}

We tested our IQT approach on the data from a 10-year-old epilepsy patient who has two cortical-subcortical cystic lesions with surrounding edema on low-field T1-weighted images at the GM-WM junction of the parietal lobes. In this case, we used IQT with ANISO U-Net(2) trained on the HCP dataset with the augmenting factor $N=1$ and the up-scaling factor of $k=4$. Figure~\ref{fig:5} shows the axial and coronal results enhanced from the low-field T1-weighted image of the patient. The IQT approach improved the GM-WM contrast globally, and significantly enhanced the resolution in coronal and sagittal planes. The enhanced image strongly highlights the two lesions in this patient which are very subtle on the input T1-weighted image. In this particular patient, the lesions were clearly visible on the original T2-weighted image, which validates that IQT highlights the lesions in the correct locations, as Fig.~\ref{fig:5}(c) shows. However, in general not all lesions are clearly visible on any MRI sequence, especially at low field, and Fig.~\ref{fig:5} highlights the potential of our algorithms to reveal subtle lesions enhancing diagnosis and potentially enabling effective treatment via clear localisation.

\section{Discussion and Conclusion}

In this work, we present an IQT approach to enhance low-field MRIs aiming to match resolution as well as contrast of high-field images. We introduce the anisotropic U-Net characterised by a deeper hierarchy and super resolving connections between input and output layers. We propose the probabilistic decimation simulator by synthesising multiple low-field images with respect to distinct grey-white matter SNR sampled from an \textit{a priori} distribution. We demonstrate that the proposed method improves the robustness on the unseen test data of variable SNR at the evaluation stage.  We validate our proposed U-Net on the evaluation dataset and the results potentially show generalisability to the actual clinical low-field images. 

This work offers several avenues for future improvement and application. Here the metrics (MSSIM and PSNR) used for quantitative assessment reflect the performance on only synthetic images. This demonstrates efficacy, but evaluation on a sizeable data set of clinical images and clinical significance from radiologists are essential for further translation. Therefore, additional qualitative evaluation by radiologist ratings and, ultimately, demonstration of improved decision making is essential to confirm impact of the approach. Nevertheless, we believe our methods have great potential to identify subtle lesions in epilepsy and other neurological conditions and thus to improve patient outcomes in LMICs in the future.

\section*{Acknowledgements}
This work was supported by EPSRC grants (EP/R014019/1, EP/R006032/1 and EP/M020533/1) and the NIHR UCLH Biomedical Research Centre. Data were provided in part by the Human Connectome Project, WU-Minn Consortium (Principal Investigators: David Van Essen and Kamil Ugurbil; 1U54MH091657) funded by NIH and Washington University. The $0.36$T MRI data were acquired at the University College Hospital, Ibadan, Nigeria.

\end{CJK*}
\end{document}